\documentclass[12pt,preprint]{aastex}
\bibpunct[, ]{(}{)}{;}{b}{}{,}
\def \aj {AJ}
\def \mnras {MNRAS}
\def \pasp {PASP}
\def \apj {ApJ}
\def \apjs {ApJS}
\def \apjl {ApJL}
\def \aap {A\&A}
\def \nat {Nature}
\def \araa {ARAA}

\newcommand{\msol} {M$_{\odot}$}

\def\lesssim{\mathrel{\hbox{\rlap{\hbox{\lower4pt\hbox{$\sim$}}}\hbox{$<$}}}}
\def\gtrsim{\mathrel{\hbox{\rlap{\hbox{\lower4pt\hbox{$\sim$}}}\hbox{$>$}}}}
\begin{document}
\title{The birth place of the type Ic Supernova 2007gr}
\shorttitle{The progenitor of SN 2007gr}
\shortauthors{Crockett et al.}

\author{R. Mark Crockett\altaffilmark{1,2}, 
	Justyn R. Maund\altaffilmark{3}, 
	Stephen J. Smartt\altaffilmark{1}, 
	Seppo Mattila\altaffilmark{1,4}, 
	Andrea Pastorello\altaffilmark{1}, 
	Jonathan Smoker\altaffilmark{1}, 
	Andrew W. Stephens\altaffilmark{5}, 
	Johan Fynbo\altaffilmark{6}, 
	John J. Eldridge\altaffilmark{1}, 
	I. John Danziger\altaffilmark{7,8}, 
	Christopher R. Benn\altaffilmark{9}}
	
\altaffiltext{1}{Astrophysics Research Centre, School of Maths and Physics, Queen's 5University Belfast, BT7 1NN, Northern Ireland, UK.}
\altaffiltext{2}{E-mail: rcrockett02@qub.ac.uk}
\altaffiltext{3}{Department of Astronomy and McDonald Observatory, University of Texas, 1 University Station, C1400, Austin, TX, 78712, U.S.A.}
\altaffiltext{4}{Tuorla Observatory, University of Turku, V\"ais\"al\"antie 20, FI-21500 Piikki\"o, Finland}
\altaffiltext{5}{Gemini Observatory, 670 North A'ohoku Place, Hilo, HI 96720, U.S.A.}
\altaffiltext{6}{Dark Cosmology Centre, Niels Bohr Institute, University of Copenhagen, Juliane Maries Vej, DK-2100 Copenhagen, Denmark.}
\altaffiltext{7}{INAF, Osservatorio Astronomico di Trieste, via G.B. Tiepolo 11, 34131 Trieste, Italy.}
\altaffiltext{8}{Department of Astronomy, University of Trieste, via G.B. Tiepolo 11, 34131 Trieste, Italy.}
\altaffiltext{9}{Isaac Newton Group, Apartado 321, E-38700 Santa Cruz de La Palma, Spain.}


\begin{abstract}

We report our attempts to locate the progenitor of the peculiar type Ic SN~2007gr in {\em HST} pre-explosion images of the host galaxy, NGC~1058. Aligning adaptive optics Altair/NIRI imaging of SN~2007gr from the Gemini (North) Telescope with the pre-explosion {\em HST} WFPC2 images, we identify the SN position on the HST frames with an accuracy of 20mas. Although nothing is detected at the SN position we show that it lies on the edge of a bright source, $134\pm23$mas (6.9\,pc) from its nominal centre. Based on its luminosity we suggest that this object is 
possibly an unresolved, compact and coeval
cluster and that the SN progenitor was a cluster member, although we note that model profile fitting favours a single bright star. 
We find two solutions for the age of this assumed
cluster; $7\mp0.5$\,Myrs and $20-30$\,Myrs, with turn-off masses of $28\pm4$\msol\ and $12-9$\msol\ respectively. Pre-explosion 
ground-based $K$-band images marginally
favour the younger cluster age/higher turn-off mass. Assuming
the SN progenitor was a cluster member, the turn-off mass provides the best estimate for its initial mass. More detailed observations, after the SN has faded, should determine if the
progenitor was indeed part of a cluster, and if so allow an age estimate to within 
$\sim$2\,Myrs thereby favouring either a high mass single
star or lower mass interacting binary progenitor.

\end{abstract}
\section{INTRODUCTION}
\label{sec:intro}

Direct detection of the progenitors of core-collapse supernovae (CCSNe) in archival observations of nearby galaxies, particularly those from the {\em Hubble Space Telescope} (HST), has recently proved successful.
Luminous red supergiant progenitors of the hydrogen rich type
II-P SNe have now been identified 
\citep{smartt03gd,2005MNRAS.364L..33M,2006ApJ...641.1060L}, 
with initial masses above
8\msol. However the progenitors of type Ib/c SN, believed to be the cores of massive stars that have lost their outer envelopes, have so far eluded
discovery \citep[e.g.][]{2005MNRAS.360..288M}. The two nearest
type Ib/c SNe with deep pre-explosion images (SN2002ap in
M74 at $\sim$9\,Mpc and SN2004gt in NGC4038 at $\sim$14\,Mpc) had no
progenitor detected \citep{2005ApJ...630L..33M,2005ApJ...630L..29G,2007arXiv0706.0500C}. The progenitors
were limited to either lower mass stars that were stripped of their hydrogen-rich envelope via binary interaction, or single
stars with masses $>30-40M_{\odot}$ for which radiatively driven mass loss may have been sufficient to remove the outer layers. In the case of the Type Ib
SN~2006jc the progenitor was discovered in an LBV-like outburst state 2 years 
before explosion, suggesting it was a very massive star 
\citep{2007Natur.447..829P}. Some fraction of Type Ib/c SNe also produce gamma-ray bursts, hence any information on their progenitors would be critical for the field \citep{2006ARA&A..44..507W}. 

In this letter we study the explosion site of SN~2007gr in NGC 1058 \cite[$10.6\pm1.3$Mpc;][]{2004A&A...425..849P} and confirm it as a type Ic. SN~2007gr was discovered by \citet{cbet1034} on 2007 Aug 15.15\,UT, located at $\mathrm{\alpha_{J2000}=2^{h}43^{m}27^{s}.98,\delta_{J2000}=+37\degr20\arcmin44\arcsec.7}$. \citet{cbet1036} spectroscopically classified SN 2007gr as Type Ib/c, but suggested that the presence of strong O I $\lambda$7774 and Ca II IR triplet absorptions were similar to those features observed in Type Ic SNe. We adopt a total reddening towards SN~2007gr of $E(B-V)=0.08$ (Valenti et al. in prep). Where standard filter sets were employed extinctions were calculated using the laws of Cardelli, Clayton \& Mathis (1989), and for HST observations using the $A_{X}/E(B-V)$ relations of \citet{1999AJ....118.2331V}.

\section{OBSERVATIONS AND DATA ANALYSIS} 
\label{sec:obs_data_analysis}

The pre-explosion site of SN~2007gr was observed on 2001 Jul 03 with
the {\em Wide Field and Planetary Camera 2} (WFPC2) on board HST as
part of programme GO-9042. Observations were made in the
F450W and F814W filters (460s each) and photometry was
carried out using the PSF-fitting photometry package
HSTphot \citep{dolphhstphot}. Further ground-based
observations of NGC1058 were found in the 4.2-m William Herschel Telescope (WHT) and the 2.5m Isaac Newton Telescope (INT)
archives. The WHT INGRID $K-$band images from 2001 Oct 6 had a total exposure time of 1320s and an image resolution of 1\farcs2. The INT WFC images from 2005 Jan 13 were taken using
$r'$-band (180s) and narrow H$_{\alpha}$ filters (600s) with image quality of 0\farcs9.

\begin{figure*}
\epsscale{1}
\plotone{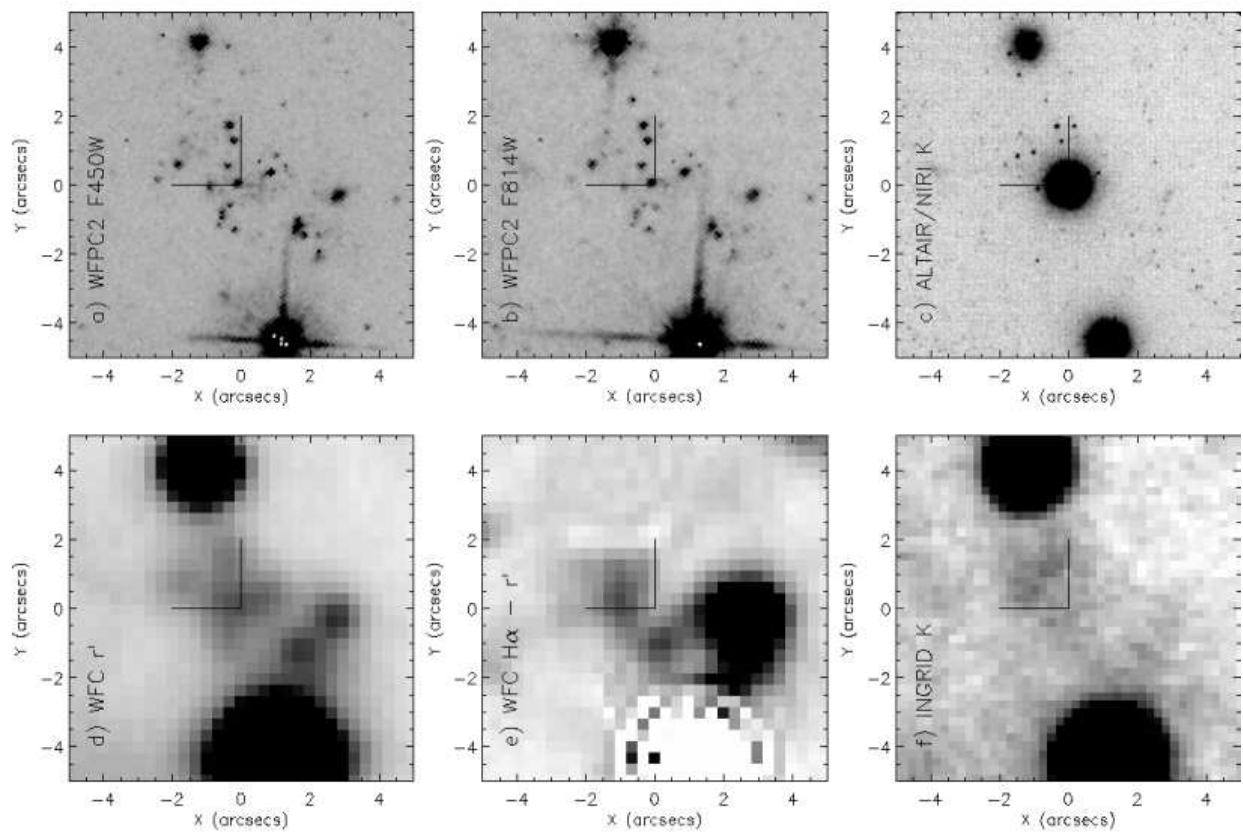}
\caption{Pre- and Post-explosion images of the site of SN 2007gr in
NGC 1058.  Each of the images is centred on the SN position and orientated such that North is up and East is to the left.  The position of SN 2007gr is located at the centre of each of the panels indicated by the cross hairs.  a) Pre-explosion HST + WFPC2 F450W image. b) Pre-explosion HST + WFPC2 F814W image. c) Post-explosion Gemini + Altair/NIRI K-band image with the SN clearly visible. d) Pre-explosion INT + WFC r'-band image. e) Pre-explosion INT + WFC $H_{\alpha}$ {\em minus} r' image. f) Pre-explosion WHT + INGRID K-band image.}
\label{fig:pre_and_post} 
\end{figure*}

Ground-based adaptive optics (AO) observations of SN~2007gr were taken
on 2007 Aug 19 using Altair/NIRI on the 8.1-m Gemini (North)
Telescope. Exposures totalling 930s were acquired 
in $K$-band using the SN as the natural guide star. The reduced image is of exceptionally high quality showing near diffraction limited resolution of 0\farcs08, with a pixel
scale of 0\farcs022 providing excellent sampling of the
PSF. Fig.\,\ref{fig:pre_and_post} shows that this Gemini image even resolves some objects which appear blended in the HST data, an
effect of the 0\farcs1 pixel scale of the WFPC2 chip. In order to confirm the classification of the SN, a spectrum was obtained on 2007 Sept 11 using the WHT equipped with ISIS (R300B+R158R).

The site of SN 2007gr fell on the WF2 chip of the
pre-explosion WFPC2 observations. To determine the position
of the SN site accurately on these images we calculated a
transformation to map the coordinate system of the post-explosion
Gemini image to the WFPC2 pre-explosion frames using the {\sc iraf geomap} task. Positions of 19 stars
common to both the WFPC2 F814W and Gemini $K$-band images provided a general
geometric transformation (involving shifts, scales
and rotation) with an RMS error of $\mathrm{\pm20}$mas. The SN position was measured in the Gemini image and transformed to the
coordinate system of the F814W pre-explosion frame yielding a pixel
position on the WF2 chip of [250.65, 118.34]. Fig.\,\ref{fig:pre_and_post}a \& b show that the SN site is very close to, but not coincident with, an object that is
bright in both the F450W and F814W images. Its position was measured as [250.10, 119.56] with an
accuracy of $\mathrm{\pm11}$mas. The SN site is
therefore 0\farcs121 west and 0\farcs056 south of this bright object.
\cite{2007CBET.1041....1L} give a position for SN2007gr on
the WFPC2 images that is around 66 mas further north
than our position, meaning our respective positions are not
coincident within the combined errors (combined error from this study $\mathrm{\pm23}$mas,
Li et al. claim $\mathrm{\pm20}$mas). We suggest that the Li et
al. error has been underestimated given that they used a MEGACAM image
of the SN from the Canada France Hawaii Telescope to perform differential astrometry, which is of much lower resolution than
our Gemini data.

The position of the SN on the INT and WHT pre-explosion observations was determined using a similar method. The INT r'-band image (Fig.\,\ref{fig:pre_and_post}d) shows significant flux at the SN position, but this is almost certainly due entirely to the resolved sources observed in the HST frames. Subtraction of the r'- band continuum from the $H_{\alpha}$ image identifies regions of strong $H_{\alpha}$ emission. Since hot OB-type stars are required to ionise the hydrogen gas, $H_{\alpha}$ emission is generally associated with clusters of young, massive stars. The image subtraction package {\sc isis} 2.2 (Alard \& Lupton 1998; Alard 2000) was used to match the PSF, flux and background of the INT r' image to its $H_{\alpha}$ counterpart. Fig.\,\ref{fig:pre_and_post}e shows several regions of quite strong emission surrounding the SN site, but none exactly coincident with its position. In the WHT K-band image (Fig.\,\ref{fig:pre_and_post}f) nothing is visible at or near the SN position.

\section{OBSERVATIONAL RESULTS AND DISCUSSION}
\label{sec:results}

SN2007gr occurred in a large, extended star-forming region that in
Fig.\ref{fig:pre_and_post} is at least 5\arcsec\ in diameter (250pc). Within this there
are bright sources, some of which are point-like and others which are
clearly extended. Most of these are likely to be compact stellar
clusters rather than single individual supergiants. Compact star clusters tend to be clustered together into 
large star forming complexes of the order 50-300\,pc, and the recent study 
of the cluster complexes in M51 shows many that are morphologically similar to the region hosting SN2007gr
\citep{2005A&A...443...79B}. HSTphot photometry of the pre-explosion WFPC2 observations gave apparent magnitudes for the bright object close to the SN site of F450W $=21.47 \pm0.02$ and F814W $=20.89 \pm0.02$. Converting these to absolute standard magnitudes using the distance to NGC~1058 and extinction corrections (Section~\ref{sec:intro}), and the colour corrections of \cite{2005MNRAS.360..288M}, gave $M_{B}=-8.99$, and $(B-I)=0.42$. If it is a single star it is a 30-40\msol~post main sequence supergiant star, or a more massive but extinguished ($A_{V}\gtrsim1$ mag) main sequence object. Such bright, single stars are not without precedent in star-forming galaxies \citep[e.g.][]{2001ApJ...548L.159B}, but they are few in number in any typical spiral due to the short lifetimes of the blue-yellow supergiant phase (e.g. Eldridge \& Tout 2004).

\begin{figure}
\epsscale{1}
\plotone{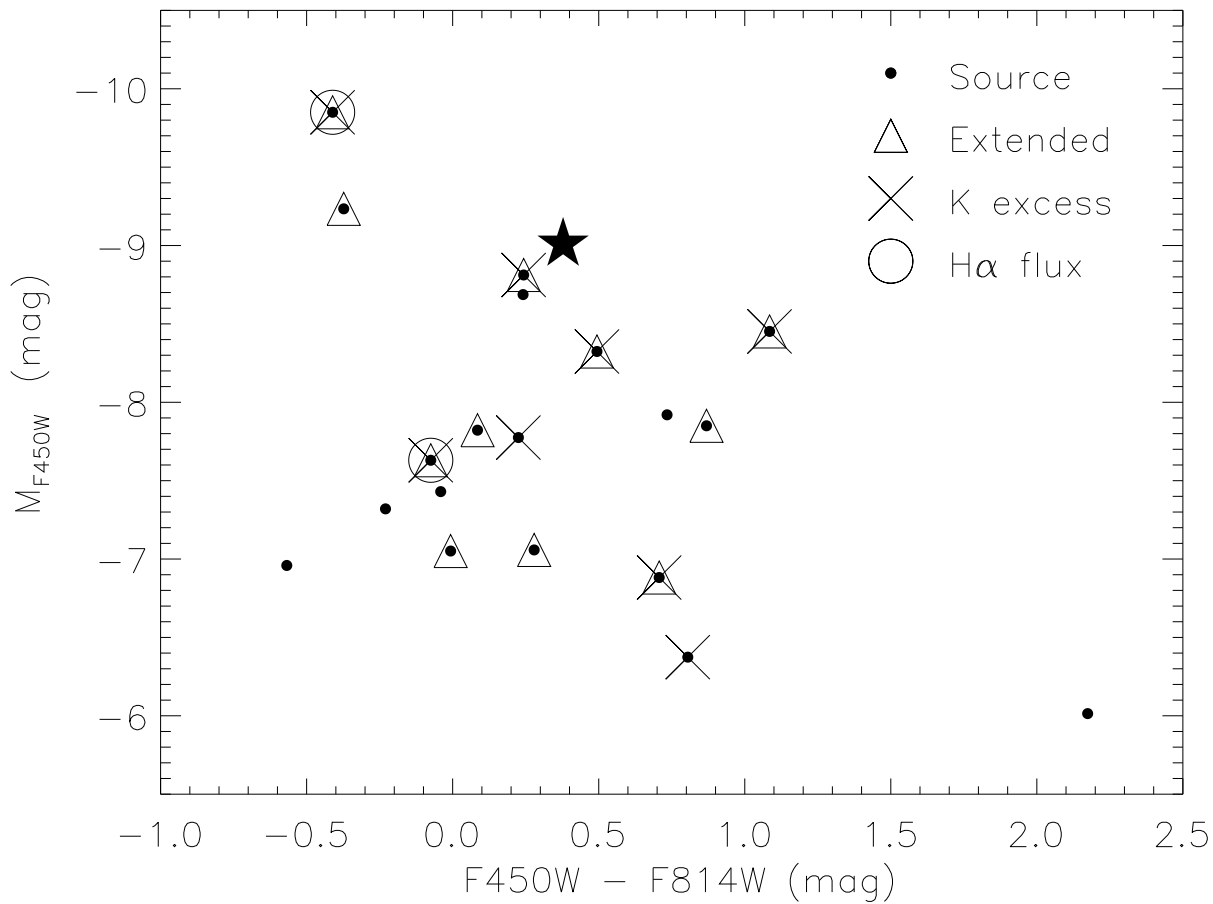}
\caption{Colour-magnitude diagram of sources within a radius of approximately $3\arcsec$ (150\,pc) of the SN position. Photometry is from the WFPC2 F450W and F814W pre-explosion observations. All sources are marked by a filled circle. Objects which {\em Ishape} measured as extended are marked with triangles, those with K band excesses of $>$0.5 mags when compared to single stellar colours are marked with crosses, and those associated with strong $H_{\alpha}$ emission are indicated by open circles. The bright object closest to the SN position is marked with a filled star. \cite{2005A&A...443...79B} suggest that anything brighter than $M_{V} <-8.6$
in these starforming complexes is likely to be an unresolved cluster rather
than a single, massive star. 
}
\label{fig:cmd} 
\end{figure}

Surrounding the SN location are several other objects of similarly high luminosity that have
colours consistent with O- to F-type supergiants
(Fig.\,\ref{fig:cmd}). It is unlikely that one would
observe such a large number of extremely massive stars during this
very short evolutionary supergiant phase in such a small surface area
of the disk. \cite{2005A&A...443...79B} suggest that sources brighter than $M_{V} <-8.6$ are most probably clusters. Several of the objects appear extended implying that they are most probably not single stars.  Most however, including the bright source nearest the SN location, are not significantly broader than the stellar PSF (FWHM$\sim$1.5 pixels (0\farcs15)), which at a distance of 10.6\,Mpc corresponds to 7.7\,pc. Given its absolute magnitudes and its proximity to the SN site ($\sim$6.9\,pc) one could suggest that this bright object is a coeval cluster of diameter less than $\sim$7.7pc, which hosted SN~2007gr. If it could be confirmed as such a cluster this is potentially a powerful way of determining the initial mass of the SN progenitor.  

In an attempt to constrain the true spatial extent of this object and others we used the {\em Ishape} routine, part of the BAOlab package (Larsen 1999, 2004). Larsen (2004) suggests that {\em Ishape} can limit the intrinsic size of a significantly detected source (S/N $\sim$50$\sigma$) down to $\sim$10 per cent of the PSF FWHM, which here corresponds to $\sim$1\,pc. Elliptical Moffat profiles with a power law index of -1.5 were convolved with TINYTIM (Krist \& Hook 1997) WFPC2 PSFs and fitted to all the objects within 3$\arcsec$ of the SN position. Those sources which are detected at $>$10$\sigma$~level and have fitted FWHM values $>$0.15 pixels (10 per cent of the PSF FWHM) in one or both of the HST observations are identified as extended in Fig.\,\ref{fig:cmd}. The bright source nearest the SN position is best fit with a delta function, suggesting that its diameter is no larger than $\sim$1\,pc, and that it is more likely a star rather than a cluster. 

However, aperture photometry of the Gemini AO image, calibrated using the two bright 2MASS sources north and south of the SN, revealed that several sources around the SN location have significant K-band excesses when compared to main sequence and supergiant stellar colours (see Fig.\,\ref{fig:cmd}), implying that they are not individual massive stars. Two of these sources are measured as point-like by {\em Ishape}. Unfortunately we cannot measure a K-band magnitude for the object nearest the progenitor site in the Gemini image due to the brilliance of the SN. Additionally the $H_{\alpha}$ image (Fig.\,\ref{fig:pre_and_post}e) gives some idea of the location of the youngest OB-stars within this large star-forming region. Sources in the HST images coincident with the $H_{\alpha}$ flux are marked in Fig.\,\ref{fig:cmd}. Their association with significant $H_{\alpha}$ emission implies that they are most likely young compact star clusters rather than single massive supergiants. We note, however, that no significant $H_{\alpha}$ emission is observed exactly coincident with the object nearest the SN site.

If this object is actually a massive single star rather than a cluster, and the progenitor is an
unrelated star within this extended star-forming region, then we can
attempt to derive a luminosity limit for the undetected SN progenitor. 
We added fake stars of various magnitudes into the WFPC2 images at the position
of the SN progenitor and tried to recover both the
real and fake sources simultaneously with PSF-fitting techniques.
We defined the detection limit in each filter as the magnitude of the fake source
that must be added to the image in order for it to be
independently detected at a position coincident with the
progenitor site. The detection limits were thereby found to be F450W $>23.7$ ($M_{\rm F450W}>-6.7$) and F814W $>21.7$ ($M_{\rm F814W}>-8.6$). The detection limit for the WHT K-band image was found to be $K>19.7$ ($M_{K}>-10.45$) by adding fake stars of increasing luminosity at the SN position and
determining, by eye, at what magnitude an obvious source became visible.

\begin{figure}
\epsscale{1}
\plotone{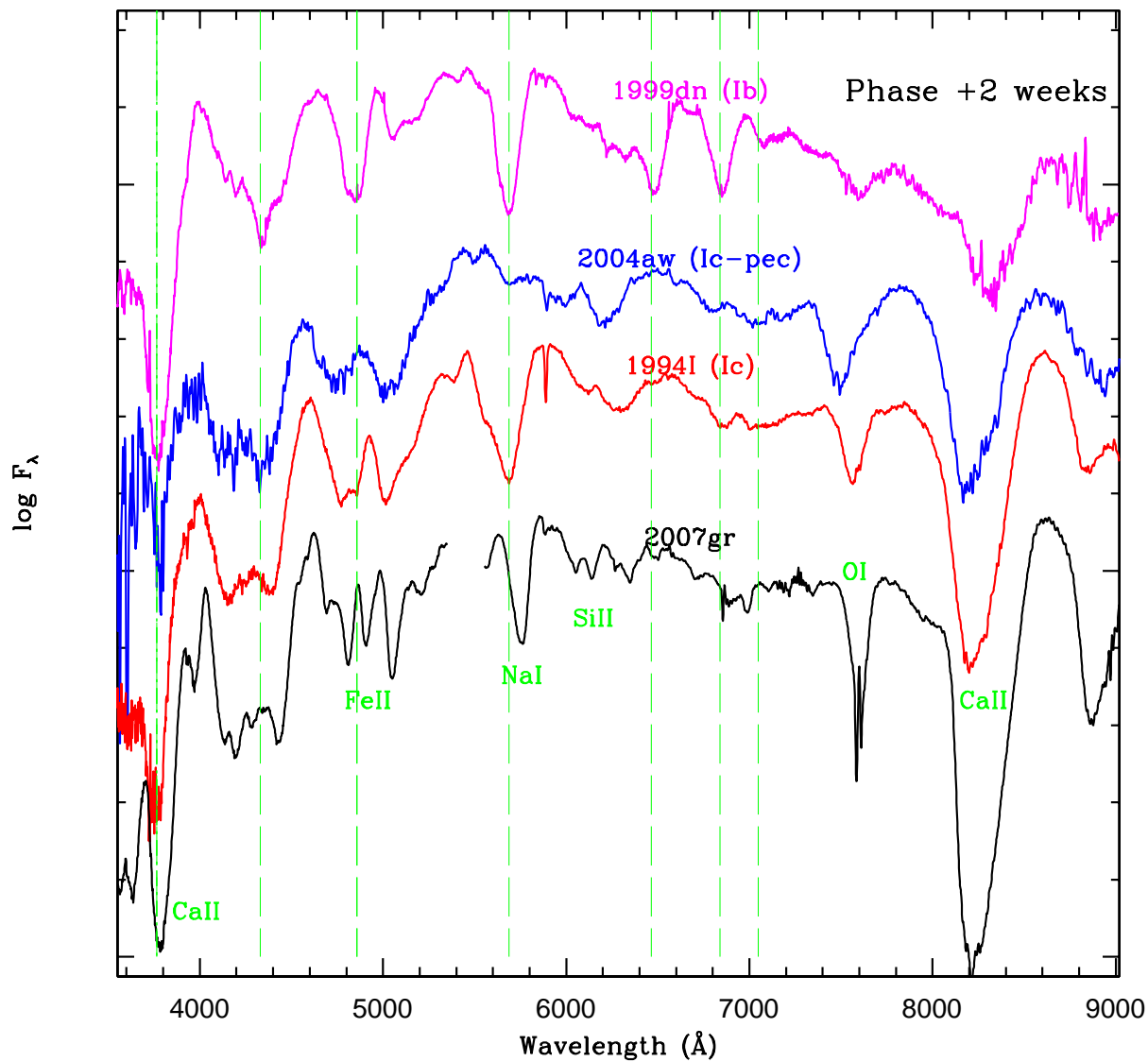}
\caption{Comparison of spectra of the classical Type Ib SN 1999dn (Branch et al.
2002), the peculiar Type Ic SN 2004aw (Taubenberger et al. 2006), the
classical SN Ic 1994I (Filippenko et al. 1995) and SN 2007gr at 12-15 days
past maximum. The spectrum of SN 2007gr shows several unusually narrow P-Cygni lines. No reddening correction was applied to the spectra, while they are reported to the host galaxy rest wavelength.
The vertical dashed lines mark the positions of the minima of the
He I lines in the spectrum of SN 1999dn, blueshifted by about 9500 km/s.
No prominent He I lines are visible in the spectrum of SN 2007gr,
suggesting a Type Ic classification. 
}
\label{fig:spectra} 
\end{figure}

Fig.\,\ref{fig:spectra} shows our WHT + ISIS spectrum compared with those of well studied type Ib/c SNe. The phase corresponds to about two weeks past maximum (Valenti et al. in prep.). While in SN~1999dn the He I lines are prominent, they are weak (if at all visible) in the type Ic SNe and also in SN~2007gr, suggesting that SN~2007gr is of type Ic. SN~2007gr shows clear similarity to SN 1994I (and, more marginally, to SN~2004aw). However, the presence of narrower features due to metals in the spectrum of SN 2007gr makes this object rather peculiar.

\section{IMPLICATIONS FOR THE PROGENITOR OF SN2007gr}
\label{sec:disc}

If we assume that the bright object
closest to the SN site is a single, massive star, one can
attempt to place constraints on the properties of the SN progenitor
using the detection limits derived in the previous section. However,
these magnitude limits are much shallower than those found for the
progenitors of SN~2002ap \citep{2007arXiv0706.0500C} and SN~2004gt
\citep{2005ApJ...630L..33M}. Since the SN was of Type Ic we can reasonably assume that its progenitor was a Wolf-Rayet (W-R) star (e.g. Eldridge \& Tout 2004). Comparing our magnitude limit of $M_{\rm F450W}>-6.7$ with B-band photometry of W-R stars shown in Fig.\,8 of \cite{2007arXiv0706.0500C}, we see that almost all of these observed W-R stars fall below our detection limit. 

If, however, the bright object closest to the SN site is an unresolved stellar cluster, then it must have a diameter of less
than  $\sim$7.7\,pc (FWHM of the PSF) or perhaps less than $\sim$1\,pc ({\em Ishape} analysis). In the case of the latter the progenitor would then be on the outskirts of the cluster, although still quite possibly coeval, e.g. NGC~3603 in the Milky Way. If SN~2007gr
arose from one of the most evolved stars in such a coeval
cluster then the cluster age would give an estimate of the age
and initial mass of the progenitor. A similar argument has been used
to estimate the progenitor mass of the type II-P SN~2004dj \citep{2004ApJ...615L.113M,2005ApJ...626L..89W}, through 
$starburst99$ population fitting of the $UBVRIJHK$ spectral energy 
distribution. The $starburst99$
code \citep{1999ApJS..123....3L} computes a coeval stellar population
at a range of ages, with a user-defined initial mass function. A cluster of mass $\sim4000$\,\msol\ at
an age of $7\mp0.5$\,Myrs produces the absolute $M_B=-8.99$ and
$B-I=0.42$ (assuming solar metallicity, an upper mass limit of 100\msol, a Salpeter IMF, and a coeval burst of
star-formation). The Geneva stellar models imply a
turn-off mass of 25\msol~for the cluster, while our STARS stellar models (Eldridge \& Tout 2004) imply a similar mass of $28\pm4$\msol. These are compatible with the minimum stellar mass thought to produce
W-R stars through single stellar
evolution \citep[e.g.][]{2001AJ....121.1050M, 2004MNRAS.353...87E}.

However this age solution is not unique as $starburst99$ can produce
$B-I \simeq 0.4$ when the cluster has evolved to $20-30$\,Myrs. At
this age the turn-off mass would be between $12-9$\msol, much too low to produce a W-R star via single stellar
evolution, and we would be forced to conclude that the progenitor was
a lower mass W-R star resulting from an 
interacting binary. This degeneracy can only be broken
with the extra wavelength coverage of the host cluster. In particular
the $U$-band is an excellent discriminator of age, and would allow its determination to within a few Myrs 
\citep{2005A&A...443...79B,2004ApJ...615L.113M}. At the older age the cluster becomes much redder in the NIR bands, with an estimated
$I-K=1.1$ and hence we would expect to detect the source at an apparent magnitude of $K=19.7$. The detection limit of the WHT K-band image lies just at this level, and this would favour the younger solution, suggesting the W-R progenitor formed from an initially 25-30\msol~star.

In approximately two years time, the SN will have faded enough that
the putative cluster can be studied in detail, in the UV, optical and NIR with
ground-based AO imaging and hopefully a fully refurbished
HST. With these higher spatial resolution and multi-wavelength images of
the immediate environment we will be able to determine if indeed it is
a compact cluster and better ascertain its properties. 
Given the difficulties in pinning down the progenitors of Ib/c SNe, even with
very deep pre-explosion imaging
\citep{2005ApJ...630L..33M,2007arXiv0706.0500C}, this particular case may offer us the best opportunity to estimate the initial mass of a type Ic progenitor. The proximity and luminosity of SN~2007gr will
make it one of the best studied, and modelled type Ic SNe to date, and it will be interesting to compare future ejecta mass estimates with our cluster turn-off mass.

\acknowledgements
Observations were from : the NASA/ESA Hubble Space Telescope obtained from the Data Archive at the Space Telescope Science Institute; the Gemini Observatory; the Isaac Newton Group Archive from CASU Astronomical Data Centre. We acknowledge funding through EURYI scheme, ESF (SJS, RMC), NSF/NASA grants AST-0406740/NNG04GL00 (JRM), Academy of Finland project: 8120503 (SM).

\clearpage

\bibliographystyle{apj}

\begin{thebibliography}{31}
\expandafter\ifx\csname natexlab\endcsname\relax\def\natexlab#1{#1}\fi


\bibitem[Alard \& Lupton (1998)]{1998ApJ...503..325A} Alard, C., Lupton, R.~H.\ 1998, \apj, 503, 325

\bibitem[Alard (2000)]{2000A&AS..144..363A} Alard, C.\ 2000, \aaps, 144, 363

\bibitem[Bastian et al.(2005)]{2005A&A...443...79B} Bastian, N., Gieles, 
M., Efremov, Y.~N., \& Lamers, H.~J.~G.~L.~M.\ 2005, \aap, 443, 79 

\bibitem[Branch et al.(2002)]{2002ApJ...566.1005B} Branch, D., et 
al.\ 2002, \apj, 566, 1005 
  
\bibitem[Bresolin et al.(2001)]{2001ApJ...548L.159B} Bresolin, F., 
Kudritzki, R.-P., Mendez, R.~H., \& Przybilla, N.\ 2001, \apjl, 548, L159

\bibitem[Bresolin et al.(2002)]{2002ApJ...567..277B} Bresolin, F., Gieren, 
W., Kudritzki, R.-P., Pietrzy{\'n}ski, G., \& Przybilla, N.\ 2002, \apj, 
567, 277 

\bibitem[Cardelli, Clayton \& Mathis (1989)]{1989ApJ...345..245C} Cardelli, J.~A., Clayton, G.~C., Mathis, J.~S.\  1989, \apj, 345, 245

\bibitem[{{Chornock} {et~al.}(2007){Chornock}, {Filippenko}, {Li} \&
{Foley}}]{cbet1036}
{Chornock}, R., {Filippenko}, A., {Li}, W. \& {Foley}, R. 2007, IAUC CBET
1036

\bibitem[{{Crockett} {et~al.}(2007){Crockett}, {Smartt}, {Eldridge},
{Mattila}, {Young}, {Pastorello}, {Maund}, {Benn} \&
{Skillen}}]{2007arXiv0706.0500C}
{Crockett}, R. M., et al.\ 2007, \mnras, 381, 835

\bibitem[{{Dolphin}(2000)}]{dolphhstphot}
{Dolphin}, A.~E. 2000, \pasp, 112, 1383

\bibitem[{{Drilling} \& {Landolt}(2000)}]{drill00}
{Drilling}, J.~S., \& {Landolt}, A.~U. 2000, in Allen's Astrophysical
  Quantities, 4th edn., ed. A.~N. {Cox} (New York: AIP)

\bibitem[{{Eldridge} \& {Tout}(2004)}]{2004MNRAS.353...87E}
{Eldridge}, J.~J., \& {Tout}, C.~A. 2004, \mnras, 353, 87

\bibitem[Filippenko et al.(1995)]{1995ApJ...450L..11F} {Filippenko}, A.~V., et 
al.\ 1995, \apjl, 450, L11 

\bibitem[Gal-Yam et al.(2005)]{2005ApJ...630L..29G} Gal-Yam, A., et al.\ 2005, \apj, 630, L29

\bibitem[Krist \& Hook (1997)]{TINYTIM} Krist, J., \& Hook, R.\ 1997, The Tiny Tim User's Guide, STScI

\bibitem[Larsen (1999)]{larsen_Ishape1} Larsen, S. S.\ 1999, A\&AS, 139, 393

\bibitem[Larsen (2004)]{larsen_Ishape2} Larsen, S. S.\ 2004, A\&A, 416, 537

\bibitem[Leitherer et al.(1999)]{1999ApJS..123....3L} Leitherer, C., et 
al.\ 1999, \apjs, 123, 3 

\bibitem[Li et al.(2006)]{2006ApJ...641.1060L} Li, W., Van Dyk, S.~D., 
Filippenko, A.~V., Cuillandre, J.-C., Jha, S., Bloom, J.~S., Riess, A.~G., 
\& Livio, M.\ 2006, \apj, 641, 1060 

\bibitem[Li et al.(2007)]{2007CBET.1041....1L} Li, W., Cuillandre, J.-C., 
van Dyk, S.~D., \& Filippenko, A.~V.\ 2007, Central Bureau Electronic 
Telegrams, 1041, 1 

\bibitem[{{Madison} \& {Li}(2007)}]{cbet1034}
{Madison}, D. \& {Li}, W. 2007, IAUC CBET 1034

\bibitem[Ma{\'{\i}}z-Apell{\'a}niz(2001)]{2001ApJ...563..151M} 
Ma{\'{\i}}z-Apell{\'a}niz, J.\ 2001, \apj, 563, 151 

\bibitem[Ma{\'{\i}}z-Apell{\'a}niz et al.(2004)]{2004ApJ...615L.113M} 
Ma{\'{\i}}z-Apell{\'a}niz, J., Bond, H.~E., Siegel, M.~H., Lipkin, Y., 
Maoz, D., Ofek, E.~O., \& Poznanski, D.\ 2004, \apjl, 615, L113 

\bibitem[Massey et al.(2001)]{2001AJ....121.1050M} Massey, P., 
DeGioia-Eastwood, K., \& Waterhouse, E.\ 2001, \aj, 121, 1050 
29

\bibitem[Matheson et al.(2001)]{2001AJ....121.1648M} Matheson, T., Filippenko, A.~V., Li, W., Leonard, D.~C., Shields, J.~C. 2001, \aj, 121, 1648

\bibitem[{{Maund} \& {Smartt}(2005)}]{2005MNRAS.360..288M}
{Maund}, J. R. \& {Smartt}, S.J. 2005, \mnras, 360, 288

\bibitem[Maund et al.(2005)]{2005MNRAS.364L..33M} Maund, J.~R., Smartt, 
S.~J., \& Danziger, I.~J.\ 2005, \mnras, 364, L33 

\bibitem[{{Maund} {et~al.}(2005){Maund}, {Smartt} \&
{Schweizer}}]{2005ApJ...630L..33M}
{Maund}, J.R., {Smartt}, S.J. \& {Schweizer}, F. 2005, \apj, 630, L33

\bibitem[Pastorello et al.(2007)]{2007Natur.447..829P} Pastorello, A., et 
al.\ 2007, \nat, 447, 8

\bibitem[Pilyugin et al.(2004)]{2004A&A...425..849P} Pilyugin, L.~S., 
V{\'{\i}}lchez, J.~M., \& Contini, T.\ 2004, \aap, 425, 849 

\bibitem[Scheepmaker et al.(2007)]{2007A&A...469..925S} Scheepmaker, R.~A., 
Haas, M.~R., Gieles, M., Bastian, N., Larsen, S.~S., \& Lamers, 
H.~J.~G.~L.~M.\ 2007, \aap, 469, 925

\bibitem[{{Schlegel} {et~al.}(1998){Schlegel}, {Finkbeiner}, \&
  {Davis}}]{schleg98}
{Schlegel}, D.~J., {Finkbeiner}, D.~P., \& {Davis}, M. 1998, \apj, 500, 525

\bibitem[{{Smartt} {et~al.}(2004){Smartt}, {Maund}, {Hendry}, {Tout},
  {Gilmore}, {Mattila}, \& {Benn}}]{smartt03gd}
{Smartt}, S.~J., {Maund}, J.~R., {Hendry}, M.~A., {Tout}, C.~A., {Gilmore},
  G.~F., {Mattila}, S., \& {Benn}, C.~R. 2004, Science, 303, 499

\bibitem[Taubenberger et al.(2006)]{2006MNRAS.371.1459T} Taubenberger, S., et 
al.\ 2006, \mnras, 371, 1459

\bibitem[Van Dyk et al.(1999)]{1999AJ....118.2331V} Van Dyk, S.~D., Peng, 
C.~Y., Barth, A.~J., \& Filippenko, A.~V.\ 1999, \aj, 118, 233

\bibitem[Wang et al.(2005)]{2005ApJ...626L..89W} Wang, X., Yang, Y., Zhang, 
T., Ma, J., Zhou, X., Li, W., Lou, Y.-Q., \& Li, Z.\ 2005, \apjl, 626, L8

\bibitem[Woosley \& Bloom(2006)]{2006ARA&A..44..507W} Woosley, S.~E., \& 
Bloom, J.~S.\ 2006, \araa, 44, 507 

\end{thebibliography}


\clearpage

\end{document}